\documentclass[%
 reprint,
superscriptaddress,
 amsmath,amssymb,
 aps,
pra,
]{revtex4-1}

\usepackage{graphicx}
\usepackage{dcolumn}
\usepackage{bm}
\usepackage{physics}
\usepackage{amsmath}

\newcommand{\bonn}{HISKP, University of Bonn, Nussallee 14-16, 53115 Bonn, Germany}
\newcommand{\bonnpi}{Physikalisches Institut, University of Bonn, Nussallee 12, 53115 Bonn, Germany}
\begin{document}

\title{Cavity-induced spin-orbit coupling in an interacting bosonic wire}

\date{\today}

\begin{abstract}
We consider theoretically ultra-cold interacting bosonic atoms confined to a wire geometry and coupled to the field of an optical cavity. A spin-orbit coupling is induced via Raman transitions employing a cavity mode and a transverse running wave pump beam, the transition imprints a spatial dependent phase onto the atomic wavefunction. Adiabatic elimination of the cavity field leads to an effective Hamiltonian for the atomic degrees of freedom, with a self-consistency condition. We map the spin-orbit coupled bosonic wire to a bosonic ladder in a magnetic field, by discretizing the spatial dimension. Using the numerical density matrix renormalization group method, we show that in the continuum limit the dynamical stabilization of a Meissner superfluid is possible, for parameters achievable by nowadays experiments. 
\end{abstract}
\author{Catalin-Mihai Halati}
\author{Ameneh Sheikhan}
\email{Corresponding author: asheikhan@uni-bonn.de}
\author{Corinna Kollath}
\affiliation{\bonn}
\affiliation{\bonnpi}
\maketitle

\section{\label{sec:Introduction}Introduction}

In the past years the experimental progress in the creation of ultracold atomic gases subjected to synthetic magnetic field or spin-orbit coupling has opened new and exciting possibilities to realize exotic quantum phases such as Meissner phases or topologically non-trivial phases in a well controlled way \cite{ZhangZhou2017, LinSpielman2009, LinSpielmanNature2009, StruckSengstock2011, MiyakeKetterle2013, AidelsburgerBloch2011, AtalaBloch2014, ManciniFallani2015, LiviFallani2016}. For cold atoms loaded into an optical lattice a pair of Raman beams can be used to induce a tunneling process between neighboring lattice sites during which the wavefunction of atoms accumulates a position dependent phase. The imprint of such a phase is due to the running wave nature of one of the Raman beams and can be interpreted as the analog of an Aharanov-Bohm phase of charged particles in a magnetic field. The spin-orbit coupling of atoms in the continuum has been realized using two-photon Raman transitions which couple internal states of the atoms for bosons \cite{LinSpielmanNature2009,LinSpielman2011} and fermions \cite{WangZhang2012, CheukZwierlein2012}. The spin-orbit coupling couples the particle's spin, here corresponding to the internal state, with its momentum, which induces chiral currents or topological effects \cite{GalitskiSpielman2013, ZhangZhang2016}. One example is a Meissner state which corresponds to a helical liquid. In this state the spin and the momentum directions of the particles couple to each other, giving rise to the propagation of the two spins in opposite directions inducing a chiral current \cite{CornfeldSela2015, OregOppen2010, LutchynSarma2010, JaparidzeMalard2014, ColeSau2017}.

The dynamic generation of gauge fields by a cavity-assisted tunneling has been proposed in recent years for cold atoms subjected to an optical lattice. The artificial magnetic field is induced dynamically via the feedback mechanism between the cavity field and the motion of atoms \cite{KollathBrennecke2016, BrenneckeKollath2016, WolffKollath2016, SheikhanKollath2016, ZhengCooper2016, BallantineKeeling2017, HalatiKollath2017}.
Phases for which the cavity mediated spin-orbit coupling plays an important role have been considered for standing-wave cavities \cite{DengYi2014, DongPu2014, PanGuo2015, PadhiGhosh2014, MivehvarRitsch2017, MivehvarPiazza2018}, or ring cavities \cite{MivehvarFeder2014, MivehvarFeder2015, OstermannMivehar2018}. 
Experimental steps took place in this direction, by realizing cavity mediated spin-dependent interactions \cite{LandiniEsslinger2018} and spinor self-ordering of bosonic atoms coupled to a cavity \cite{KroezeLev2018}.
Theoretically, the steady state diagram has been determined for a ladder geometry in the case of noninteracting fermions  \cite{BrenneckeKollath2016, KollathBrennecke2016, WolffKollath2016} and interacting bosons \cite{HalatiKollath2017}, where states with finite chiral currents have been found, or non-trivial topological properties in two dimensions \cite{SheikhanKollath2016}.

In the present work we consider interacting bosons confined to a one-dimensional wire with a dynamically induced spin-orbit coupling via the coupling to a cavity mode. We consider a set of parameters close to the ones of existing experimental setups and characterize the self-organized states that arise. We show the dynamic stabilization of a state with a persistent chiral spin current, the Meissner superfluid, in the coupled atomic cavity system. 

In Sec.~\ref{sec:setup} we describe the setup of the interacting bosonic wire placed into the optical cavity and the theoretical model. In Sec.~\ref{sec:adiabiatic} we derive an effective model for the atomic degrees of freedom and a stability condition by performing the adiabatic elimination of the cavity field. In Sec.~\ref{sec:lattice} we introduce the discretization which we use in order to simulate the system on a lattice. In Sec.~\ref{sec:obs} we present the observables used for the characterization of the Meissner superfluid state. In Sec.~\ref{sec:method} the typical parameters used within the numerical density matrix renormalization group method are given. The dynamical stabilization of the Meissner superfluid as a self-organized state with a finite cavity occupation is presented in Sec.~\ref{sec:results}. 

\section{\label{sec:Model}Model}

\subsection{\label{sec:setup}Description of the setup}

\begin{figure}[hbtp]
\centering
\includegraphics[width=.5\textwidth]{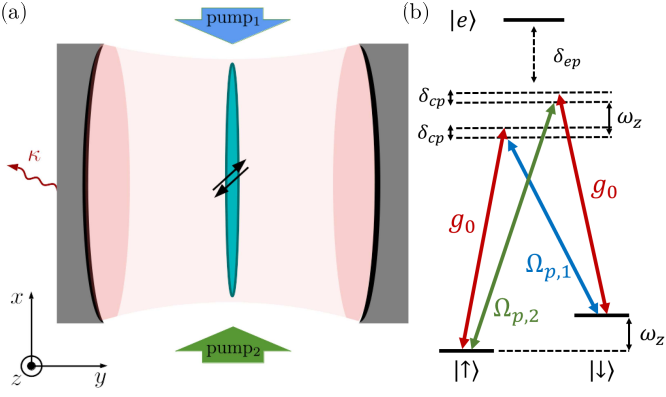}
\caption{(a) Sketch of the setup. The bosonic atoms in an optical cavity are confined in a one-dimensional wire. (b) Level scheme of the cavity-induced Raman coupling: $\ket{\uparrow}$ and $\ket{\downarrow}$ denote two internal states, and $\ket{e}$ the excited internal electronic state. The energy offset between the two states is $\hbar\omega_z$. The spin-orbit coupling is realized by two Raman processes each of which involve the cavity mode with vacuum Rabi frequency $g_0$ and one of the two transverse running wave pump beam with Rabi frequency $\Omega_{p,i=1,2}$. $\delta_{ep}$ and $\delta_{cp}$ are the frequencies of the excited state and the cavity mode in the rotating frame. 
 }
\label{fig:setup}
\end{figure}

We study an ultracold bosonic gas placed in an optical cavity confined to a one-dimensional wire (Fig.~\ref{fig:setup}). To create the spin-orbit coupling we use three internal states of an atom, as depicted in Fig.~\ref{fig:setup}(b). The states $\ket{\uparrow}$ and $\ket{\downarrow}$ differ in energy by $\hbar\omega_z$, and can be coupled using two balanced Raman transitions, each one involving a transverse running-wave pump laser and a standing-wave cavity mode (Fig.~\ref{fig:setup}). The transverse pump laser beams have the frequencies $\omega_{p,i=1,2}$, the Rabi frequencies $\Omega_{p,i=1,2}$ and the wave-vectors ${\bf k}_{p,i}=k_{p}{\bf e}_x$, along the $x$-direction, where the unit vectors along the three spatial directions are $\{{\bf e}_x,{\bf e}_y,{\bf e}_z\}$. The difference between the two pump beam frequencies is $\hbar(\omega_{p,2}-\omega_{p,1})\approx 2\hbar \omega_z$. The cavity mode has the frequency $\omega_c$, vacuum Rabi frequency $g_0$ and the wave-vector ${\bf k}_c=k_c{\bf e}_y$. We assume that all the other cavity modes are far detuned from the possible transitions. The detuning between the cavity mode and the first pump beam is tuned such that it is close to the offset, $\hbar(\omega_c-\omega_{p,1})\approx\hbar\omega_z$, and the cavity and pump modes are considered to be far detuned from the internal atomic transition to the excited state, i.e. $\omega_e\gg\omega_c,\omega_{p,i=1,2}$, thus, the excited state population is negligible and can be adiabatically eliminated. In the following we will use the rotating frame with the frequency $\omega_p=(\omega_{p,2}+\omega_{p,1})/2$.

The spin-orbit coupling is obtained via the cavity-induced Raman tunneling.
During the Raman transition a spatially dependent phase factor $\text{e}^{-i \Delta\bf{k\cdot r}}$ is imprinted onto the atomic wavefunction, where $\Delta{\bf k}=k_{p}{\bf e}_x\pm k_c{\bf e}_y$ and ${\bf r} =x {\bf e}_x$. This corresponds to a dynamically induced spin-orbit coupling. As the cavity mode does not give a contribution, the imprinted flux is $\varphi=k_{p}$, defining a corresponding length-scale $l_\varphi\equiv 2\pi/\varphi$. The strength of the spin-orbit coupling can be varied, in an experimental realization, by tilting the pump beams away from the ${\bf e}_x$ direction.

The effective description for the atoms after the adiabatic elimination of the excited state, derived similarly to Refs.~\cite{RitschEsslinger2013, NagyDomokos2008}, reads

\begin{align}
\label{eq:Hamiltonian}
&H=H_c+H_{kin}+H_{int}+H_{trap}+H_{ac} \\
&H_c= \hbar\delta_{cp} a^\dagger a\nonumber\\
&H_{kin}=-\frac{\hbar^2}{2 m} \sum_{\sigma=\uparrow,\downarrow}\int dx~\psi_{\sigma}^\dagger(x) \partial_x^2\psi_{\sigma}(x)\nonumber\\
&H_{int}=\frac{U}{2} \sum_{\sigma=\uparrow,\downarrow} \int dx~\psi_{\sigma}^\dagger(x)^2\psi_{\sigma}(x)^2   \nonumber\\
&\quad\quad +V\int dx~\psi_{\uparrow}^\dagger(x)\psi_{\uparrow}(x)\psi_{\downarrow}^\dagger(x)\psi_{\downarrow}(x)\nonumber\\
&H_{trap}=\frac{4V_{trap}}{\mathcal{L}^2} \sum_{\sigma=\uparrow,\downarrow}\int dx~(x-x_0)^2 \psi_{\sigma}^\dagger(x)\psi_{\sigma}(x) \nonumber\\
&H_{ac}=  -\hbar\tilde{\Omega} ( a + a^\dagger) ( K_{\uparrow\downarrow} + K_{\uparrow\downarrow}^\dagger)\nonumber\\
&K_{\uparrow\downarrow}=  \int dx~e^{ix\varphi }\psi_{\uparrow}^\dagger(x) \psi_{\downarrow}(x)\nonumber.
\end{align}

$H_c$ describes the cavity mode in the rotating frame, where $\delta_{cp}=\omega_c-\omega_p$ and $a$ and $a^\dagger$ are the bosonic annihilation and creation operators for the cavity mode. The bosonic field operators $\psi_\sigma(x)$ and $\psi_\sigma^\dagger(x)$ are the annihilation and creation operators of the atoms in the wire for the spin state $\sigma=\uparrow,\downarrow$. The length of the one-dimensional system is denoted by $\mathcal{L}$. The total number of bosonic atoms is $N$, with the mean inter-particle spacing $d=\mathcal{L}/N$ and the density of the wire $\rho=d^{-1}$. $H_{kin}$ describes the kinetic term for the atoms along the direction of the wire, with $m$ the mass of the atoms. The repulsive contact interaction is given in $H_{int}$, where the strength of the interaction of the atoms with the same spin is spin independent and denoted by $U$ and between atoms with different spin is $V$, ($U,V>0$). In nowadays experimental setups an external harmonic trapping potential is often present. We included this effect by adding the term $H_{trap}$, with $x_0=\mathcal{L}/2$. $H_{ac}$ describes the coupling between the atoms and the cavity field. The creation or annihilation of a cavity photon is accompanied by a spin flip. By the spatial dependence of the imprinted phase during the spin flip a dynamically induced spin-orbit coupling for the atoms is realized. In order to prevent a privileged spin state we couple the spin flip in each direction to both the creation and the annihilation operators of the cavity field, by using two pump laser beams \cite{DimerCarmichael2007}. The strength of this process is given by the amplitude $\tilde{\Omega}=\frac{\Omega_{p,1}g_0}{\omega_e-\omega_{p,1}}$, where $g_0$ is the Rabi frequency of the cavity mode. We fix the Rabi frequency of the second pump beam to $\Omega_{p,2}= \Omega_{p,1}\frac{\omega_e-\omega_{p,2}}{\omega_e-\omega_{p,1}}$, in order to realize the balanced Raman scheme. 

We describe the dynamics of the coupled cavity atom system by a Lindblad equation. 
The Lindblad description is used in order to take the cavity losses into account in addition to the unitary evolution.
The losses are due to the imperfections of the cavity mirrors. The evolution of an arbitrary operator $O$ is given by

\begin{align}
\label{eq:Lindblad}
& \pdv{t} O = \frac{i}{\hbar} \left[ H, O \right] + \mathcal{D}(O), 
\end{align}
with the dissipator $\mathcal{D}(O) =  \kappa \left( 2a^\dagger O a - O a^\dagger a - a^\dagger a O  \right)$, which describes the loss of cavity photons.

\subsection{\label{sec:adiabiatic}Adiabatic elimination of the cavity field}

In the following, employing the adiabatic elimination of the cavity field \cite{RitschEsslinger2013, KollathBrennecke2016, HalatiKollath2017}, we derive an effective model for the bosonic atoms. Within this approximation, the cavity field is replaced with its steady state value, computed from the condition $\partial_t \langle a\rangle=0$. Using Eq. (\ref{eq:Lindblad}) this condition becomes

\begin{equation}\label{eq:dyn_photon}
i \partial_t \langle a\rangle=-\tilde{\Omega}\langle K_{\uparrow\downarrow}+K_{\uparrow\downarrow}^\dagger \rangle +(\delta_{cp}- i \kappa ) \langle a\rangle=0.
\end{equation}

This relates the expectation value of the cavity field to the expectation value of $K_{\uparrow\downarrow}$ by
\begin{equation}\label{eq:alpha}
\langle a \rangle = \frac{\tilde{\Omega}}{\delta_{cp}- i \kappa }\langle  K_{\uparrow\downarrow}+K_{\uparrow\downarrow}^\dagger\rangle.
\end{equation} 

The model exhibits a $\mathbb{Z}_2$ symmetry, associated with the inversion of the sign of both the cavity field, $a+a^\dagger$, and of $K_{\uparrow\downarrow}+K_{\uparrow\downarrow}^\dagger$.  We choose without loss of generality $\langle K_{\uparrow\downarrow}+K_{\uparrow\downarrow}^\dagger\rangle>0$.

By performing a mean-field decoupling of the atomic and cavity degrees of freedom, we derive the following equations of motion for the atomic operators

\begin{align}
\label{eq:dyn_boson}
i \hbar \partial_t &\langle \psi_\sigma(x) \rangle=-\frac{\hbar^2}{2 m}\partial_x^2\langle \psi_\sigma(x) \rangle+U\langle \psi_\sigma^\dagger(x)\psi_\sigma(x)\psi_\sigma(x) \rangle\nonumber \\
&+V\langle \psi_\sigma(x)\psi_{\overline{\sigma}}^\dagger(x)\psi_{\overline{\sigma}}(x) \rangle-\hbar\tilde{\Omega}\langle a+a^\dagger\rangle e^{\pm i\varphi x} \langle \psi_{\overline{\sigma}}(x) \rangle \nonumber\\
&+\frac{4V_{trap}}{\mathcal{L}}(x-x_0)^2 \langle \psi_\sigma(x) \rangle,
\end{align}
with $\sigma=\uparrow \text{or} \downarrow$ and the sign in the exponential is positive for the spin state $\sigma=\uparrow$ and negative for $\sigma=\downarrow$ .

The dynamics given by substituting the expectation value of the cavity field, Eq.~(\ref{eq:alpha}), into the equations of motion of the bosonic operators, Eq.~(\ref{eq:dyn_boson}), can be effectively described by the following Hamiltonian

\begin{align}
\label{eq:Hamiltonian_eff}
&H_{eff}=H_{kin}+H_{\uparrow\downarrow}+H_{int}+H_{trap} \\
&H_{\uparrow\downarrow}=-J_{\uparrow\downarrow} (K_{\uparrow\downarrow} + K_{\uparrow\downarrow}^\dagger).\nonumber
\end{align}
In the effective Hamiltonian, Eq.~(\ref{eq:Hamiltonian_eff}), $J_{\uparrow\downarrow}$ gives the amplitude of the spin-orbit coupling. $J_{\uparrow\downarrow}$ has to be computed self-consistently because it depends on the occupation of the cavity field and thus on the expectation value of $\langle K_{\uparrow\downarrow} \rangle$. The self consistency condition is given by
\begin{equation}
\label{eq:sc_cond}
J_{\uparrow\downarrow}=A\langle K_{\uparrow\downarrow} \rangle, \text{ with } A= \frac{4\hbar\tilde{\Omega}^2\delta_{cp}}{\delta_{cp}^2 +\kappa^2}.
\end{equation}
We will call $A$, loosely, the pump strength, since this is one of the experimental knobs used to tune $A$. Depending on the parameters non-trivial self consistent solutions can be typically found for a continuous range of values of $A$.

The stability of the non-trivial solutions, obtained as the ground states of the effective model, needs to be investigated. A stability condition can be derived by considering perturbations around the steady state \cite{RitschEsslinger2013,HalatiKollath2017,Tian2016}. 
We follow an analogous stability analysis as performed in Ref.~\cite{HalatiKollath2017}. By looking at the behavior of linear fluctuations around the stationary solutions of the equations of motion for the cavity field, Eq.~(\ref{eq:dyn_photon}), we get the following stability condition

\begin{equation}
\label{eq:condition}
\frac{d\langle K_{\uparrow\downarrow} \rangle}{dJ_{\uparrow\downarrow}}<\frac{1}{A}.
\end{equation}

\subsection{\label{sec:lattice}Discrete lattice model}

In order to determine the self-consistent solutions we discretise the spatial dimension with a spacing $\Delta x$. This maps the effective Hamiltonian, Eq.~(\ref{eq:Hamiltonian_eff}), to the corresponding model on a lattice. The derived lattice Hamiltonian reads

\begin{align}
&H_{BH}=H_{kin}+H_{\uparrow\downarrow}+H_{int}+H_{trap} \label{eq:discrete_ham}
\\
&H_{kin}=-\frac{J}{\Delta x^2}\sum_{l,\sigma=\uparrow,\downarrow} (b_{\sigma,l}^\dagger b_{\sigma,l+1}^{\phantom\dagger} + b_{\sigma,l+1}^\dagger b_{\sigma,l}^{\phantom\dagger})\nonumber\\
&H_{\uparrow\downarrow}= -J_{\uparrow\downarrow} (K_{\uparrow\downarrow} + K_{\uparrow\downarrow} ^\dagger) \nonumber\\
&K_{\uparrow\downarrow}=  \sum_{l} e^{i\varphi \Delta x l}b_{\uparrow,l}^\dagger b_{\downarrow,l}^{\phantom\dagger}\nonumber \\
&H_{int}=\frac{U}{2\Delta x} \sum_{l,\sigma=\uparrow,\downarrow} n_{\sigma,l}(n_{\sigma,l}-1)+\frac{V}{\Delta x}\sum_{l}n_{\uparrow,l}n_{\downarrow,l}\nonumber\\
&H_{trap}=\frac{4 V_{trap}}{\mathcal{L}^2}\Delta x^2\sum_{l,\sigma=\uparrow,\downarrow}\qty(l-l_0)^2n_{\sigma,l}.\nonumber
\end{align}
The bosonic operators $b_{\sigma,l}$ and $b_{\sigma,l}^\dagger$ are the annihilation and creation operators of the atoms where $\sigma=\uparrow,\downarrow$ labels the legs of the ladder and $l=1,...,L$ the rungs of the ladder. $L$ denotes the number of the rungs of the ladder and it is related to the physical size of the system by $L=\frac{\mathcal{L}}{\Delta x}+1$. The operator $n_{\sigma,l}=b_{\sigma,l}^\dagger b_{\sigma,l}^{\phantom\dagger}$ is the number operator and $J=\frac{\hbar^2}{2m}$. $H_{BH}$ corresponds to a Bose-Hubbard ladder in a magnetic field, where the two spins states represent the two legs of the ladder and spin flip processes are equivalent to the tunneling along the rungs of the ladder. Compared to our previous work from Ref.~\cite{HalatiKollath2017} on the bosonic ladder in a magnetic field, here we have additionally a non-local interaction along the rungs of the ladder, given by $V$. This model has been studied previously in Refs.~\cite{OrignacGiamarchi2001, PetrescuHur2015, Uchino2016, GreschnerVekua2016, StrinatiMazza2017}. The parameters of the ladder model are renormalized such that in the continuum limit, $\Delta x\to 0$, we would recover the Hamiltonian given by Eq.~(\ref{eq:Hamiltonian_eff}). 

The procedure of determining the stable steady states of the model, Eq.~(\ref{eq:Hamiltonian}), consists in four steps: 
\begin{enumerate}
\item First, we perform ground-state searches for the effective discrete model, Eq.~(\ref{eq:discrete_ham}), using a matrix product state method (MPS), for a fixed discrete spacing $\Delta x$ and physical parameters of the continuum model: system size $\mathcal{L}$, particle number $N$, magnetic flux $\phi$, and interactions $U/J$ and $V/J$, while varying $J_{\uparrow\downarrow}/J$. 
\item We compute the expectation value $\langle K_{\uparrow\downarrow} \rangle$ as a function of $J_{\uparrow\downarrow}/J$ and solve the self-consistency equation. The self-consistency condition, Eq.~(\ref{eq:sc_cond}), can be interpreted graphically, using the reformulation

\begin{equation}
\label{eq:condition2}
\frac{\langle K_{\uparrow\downarrow} \rangle}{\mathcal{L}d^{-1}}=\frac{J}{A\mathcal{L}d}~\qty(\frac{J_{\uparrow\downarrow}}{Jd^{-2}}).
\end{equation}

The self-consistent solution correspond to the crossing of the two curves. The slope of the RHS depends on the pump strength $A$. Thus, the solutions have to be determined for each value of $A$.
\item The stability of the non-trivial solutions is inferred by comparing the slopes of the left-hand and right-hand sides of Eq.~(\ref{eq:condition2}), we see from the stability condition, Eq.~(\ref{eq:condition}), that a solution is stable if the slope of  $\langle K_{\uparrow\downarrow} \rangle$ is smaller than $\frac{J}{A\mathcal{L}d}$. 
\item The last step is to verify the convergence of the results in the continuum limit  $\Delta x \to 0$ by considering different values of $\Delta x$.
\end{enumerate}

\subsection{\label{sec:obs}Observables in the Meissner phase}

In this section we will describe the observables that we use to characterize the non-trivial stable steady states of the system and how we can identify them with the corresponding observables computed on the ladder. We will focus on the properties of the Meissner superfluid, as this can be dynamically stabilized in the cavity, for the parameters considered in this work. The Meissner state is a "zero momentum" state, as close to the non-interacting regime the bosonic quasiparticles are condensed in the single-particle dispersion minimum at momentum $k=0$. It corresponds to a helical liquid, where the spin and the momentum directions of the particles lock to each other and with the two spins propagating in opposite directions \cite{CornfeldSela2015, OregOppen2010, LutchynSarma2010, JaparidzeMalard2014, ColeSau2017}, giving rise to a chiral current.

We start with the local densities and currents, as their configurations can point towards the Meissner or vortex nature of the chiral phases. The local density $n_\sigma(x)$, the local current $j_{\sigma}(x)$, and the spin flip current $j^{\uparrow\downarrow}(x)$ are defined and computed as

\begin{align}
\label{eq:localcur}
n_\sigma(x) &= \psi_\sigma^\dagger(x)\psi_\sigma^{\phantom\dagger}(x) \approx \frac{1}{\Delta x } b_{\sigma,l}^\dagger b_{\sigma,l}^{\phantom\dagger} \nonumber \\
j_{\sigma}(x) &= -i J\qty[\psi_\sigma^\dagger(x)\partial_x\psi_\sigma^{\phantom\dagger}(x)-(\partial_x\psi_\sigma^\dagger(x))\psi_\sigma^{\phantom\dagger}(x)]\nonumber\\
&\approx -i \frac{J}{\Delta x^2}\qty(b_{\sigma,l}^\dagger b_{\sigma,l+1}^{\phantom\dagger} -\text{H.c.}),\nonumber \\
j^{\uparrow\downarrow}(x) &=-i J_{\uparrow\downarrow}  \qty(e^{i\varphi x}\psi_{\downarrow}^\dagger(x) \psi_{\uparrow}^{\phantom\dagger}(x) -\text{H.c.}) \nonumber\\
&\approx -i\frac{J_{\uparrow\downarrow}}{\Delta x} \qty(e^{i\varphi \Delta x l}b_{\downarrow,l}^\dagger b_{\uparrow,l}^{\phantom\dagger} -\text{H.c.}), 
\end{align}
with $x=l\Delta x $.
From the local currents, one can compute global observables as the chiral current $J_c$, which is a persistent spin current

\begin{align}
\label{eq:cur}
&J_c = \frac{1}{\mathcal{L}} \int dx~ \langle j_{\uparrow}(x) - j_{\downarrow}(x) \rangle.
\end{align}

The Meissner phase is characterized in an infinite homogeneous system by zero spin flip currents in the bulk and a finite value of the chiral current. The values of the chiral current and of the expectation value of $\langle K_{\uparrow\downarrow} \rangle$ can be computed by approximating the integrals with sums, $\int dx ~O(x)\approx \Delta x \sum_l ~O(l\Delta x)$. 

In order to confirm the superfluid nature of the states we can use the single particle correlation and look for the algebraic decay, a characteristic feature of the standard Luttinger liquid paradigm \cite{Giamarchibook}. Here we use $\langle \psi^\dagger_{\sigma}(x_0) \psi_{\sigma}(x_0+x)^{\phantom\dagger}+\text{H.c.} \rangle \approx\langle b^\dagger_{\sigma,l_0} b_{\sigma,l_0+l}+\text{H.c.} \rangle\propto l^{-\alpha}$. 
In the Meissner superfluid the central charge $c$, which can be interpreted as the number of gapless modes, is $c=1$, as this phase has a gapless symmetric sector due to the superfluid nature of the state and a gapped antisymmetric sector. The central charge can be extracted from the scaling of the von Neumann entropy $S_{vN}(l)$ of an embedded subsystem of length $l$ in a system of length $L$. We compute $S_{vN}(l)$ for a bipartition into two subsystems of length $l$ and $L-l$. For open boundary conditions the von Neumann entropy for the ground state of gapless phases scales as \cite{VidalKitaev2003,CalabreseCardy2004, HolzeyWilczek1994}

\begin{equation}
\label{eq:entropy}
S_{vN}=\frac{c}{6}\ln\left(\frac{L}{\pi}\sin\frac{\pi l}{L}\right)+s_1,
\end{equation}
where $s_1$ is a non-universal constant and we have neglected oscillatory terms \cite{LaflorencieAffleck2006} due to the finite size of the system and logarithmic corrections \cite{AffleckLudwig1991}. 

In the numerical simulations, due to the lattice discretization and the open boundary conditions, we observe oscillations in the local density and currents, which are algebraically decaying away from the boundaries. In order to reduce the influence of these boundary effects on the computed observables, we calculate $J_c$ and $\langle K_{\uparrow\downarrow} \rangle$ from the average around which the oscillations occur, extracted by fitting the density and currents in the center of the system, with the function $f(x)=a+b\cos(cx+\phi)$. We show that in the Meissner phase the amplitude of these oscillations decreases in the continuum limit and vanishes in the thermodynamic limit.

\section{\label{sec:method}Method}

The numerical results presented in this work were obtained using a finite-size density matrix renormalization group (DMRG) algorithm in the matrix product state form \cite{White1992, Schollwock2005, Schollwock2011, Hallberg2006, Jeckelmann2008}, using the ITensor Library \cite{itensor}. We simulate the discrete model Eq.~(\ref{eq:discrete_ham}) for ladders with a number of rungs between $L=125$ and $L=275$, depending on the chosen discretisation $\Delta x$, and with a bond dimension up to 1500. The convergence of the method to a sufficient accuracy was assured by comparing results with different truncation errors and bond dimensions. Since we are considering a bosonic model with finite interactions the local Hilbert space is infinite, thus, a cutoff for its dimension is needed. We use a maximal local dimension of three bosons per site, as we are dealing with a low density of particles on the ladder. The higher cutoff of four bosons per site gives consistent results.

\section{\label{sec:results}Results}

\subsection{\label{sec:stable}Identifying the stable stationary states in the homogeneous system}

\begin{figure}[hbtp]
\centering
\includegraphics[width=.5\textwidth]{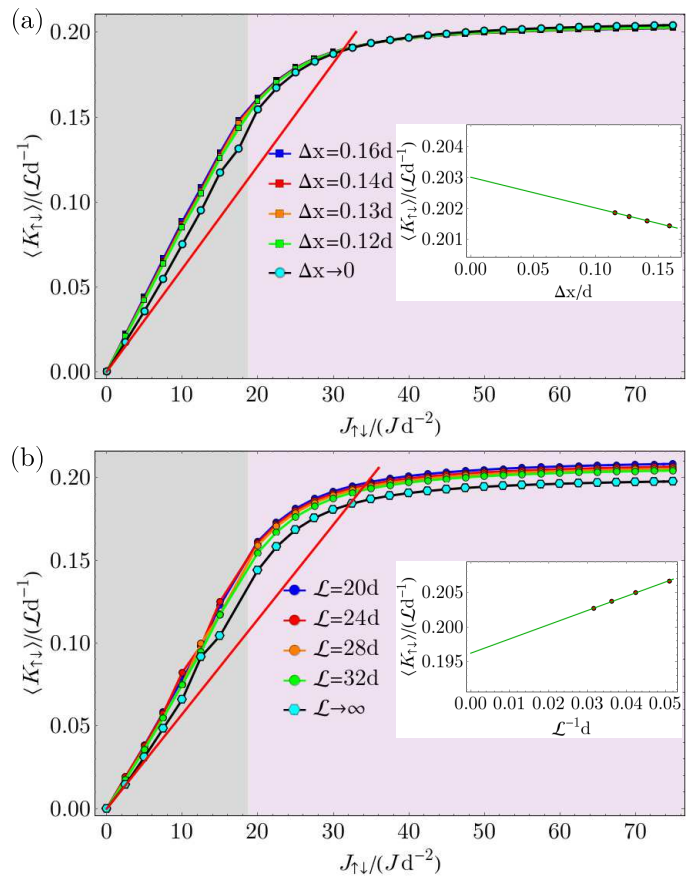}
\caption{Graphical interpretation of the self-consistency condition for the parameters $l_\varphi=0.8d$, $U=32.5Jd^{-1}$, $V=30Jd^{-1}$ and $V_{trap}=0$. The expectation value $\langle K_{\uparrow\downarrow} \rangle/Ld^{-1}$ is represented for (a) $\mathcal{L}=32d$ and multiple values of $\Delta x$, together with an extrapolation to the limit $\Delta x\to 0$, (b) for multiple system sizes in the continuum limit, and the extrapolation to the thermodynamic limit $1/\mathcal{L}\to 0$.
The straight (red) lines represent the right-hand side of the self-consistency condition, Eq.~(\ref{eq:condition2}), which is a linear function with slope $\frac{J}{A\mathcal{L}d}$, for one chosen value of $A$. 
The intersections of the two curves give the solutions of the self-consistency condition.
In the insets the extrapolations (a) $\Delta x \to 0$ and (b) $1/\mathcal{L}\to 0$ of $\langle K_{\uparrow\downarrow} \rangle/\mathcal{L}d^{-1}$, for $J_{\uparrow\downarrow}=62.5Jd^{-2}$, are depicted. 
The purple shaded area marks the regime in which for a certain value of $A$ a stable solution can be found  and the grey shaded area the regime in which our results are not accurate enough in order to determine the stability conclusively.
 }
\label{fig:kpersol79}
\end{figure}

\begin{figure}[hbtp]
\centering
\includegraphics[width=.5\textwidth]{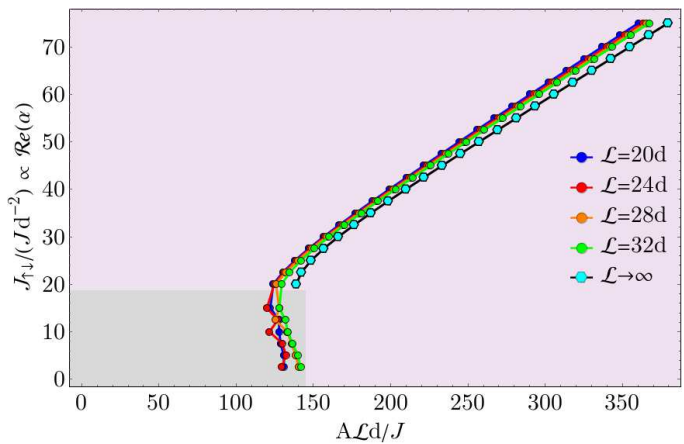}
\caption{ The solutions $J_{\uparrow\downarrow}/Jd^{-2}$ of the self-consistency equation  which are proportional to the cavity field $ Re(\alpha)$ versus the pump strength $A\mathcal{L}d/J$, for multiple $\mathcal{L}$, in the continuum and thermodynamic limits. In the grey area the stability of the solutions is not clear.
 }
\label{fig:kpersol}
\end{figure}

In this subsection we solve the self-consistency condition and identify the steady states which can be stabilized for different values of $J_{\uparrow\downarrow}$, for magnetic length $l_\varphi=0.8d$, interaction $U=32.5Jd^{-1}$ and $V=30Jd^{-1}$ in a homogeneous system, $V_{trap}=0$. We show that the dynamic stabilization of a Meissner superfluid state is possible.
We calculate the expectation value $\langle K_{\uparrow\downarrow} \rangle/\mathcal{L}d^{-1}$ in the ground states of the effective model. The self-consistent solutions are given by intersections of this curve with the linear function  $\frac{J}{A\mathcal{L}d}~\qty(\frac{J_{\uparrow\downarrow}}{Jd^{-2}})$, as we can see in Fig.~\ref{fig:kpersol79}(a) for $\mathcal{L}=32d$ and different values of $\Delta x$. For the stability of the solutions we compare the slopes of the two curves, Eq.~(\ref{eq:condition2}). If the derivative of  $\langle K_{\uparrow\downarrow} \rangle/\mathcal{L}d^{-1}$ is less than the slope of the linear function, the solution is stable. 

For the considered parameters (Fig.~\ref{fig:kpersol79}), we can find stable solutions for $J_{\uparrow\downarrow}\gtrsim 18.75Jd^{-2}$, for all considered values of $\Delta x$. 

As the values of $\Delta x$ are finite, these results indicate that the stabilization of the Meissner superfluid state is possible in a system which has an underlying lattice potential. In order to estimate the value of $\langle K_{\uparrow\downarrow} \rangle/\mathcal{L}d^{-1}$ in the limit $\Delta x \to 0$, we perform an extrapolation of our data for finite $\Delta x$, as shown in the inset of Fig.~\ref{fig:kpersol79}(a) for $J_{\uparrow\downarrow}=62.5Jd^{-2}$. The value of $\langle K_{\uparrow\downarrow} \rangle/\mathcal{L}d^{-1}$ in the continuum limit is labeled by $\Delta x\to 0$ in Fig.~\ref{fig:kpersol79}(a). Also the extrapolation $\Delta x \to 0$ leads to a stable solution, such that we are confident that the solutions remain stable in the continuum limit. 

In contrast for small values of $J_{\uparrow\downarrow}$, marked by the grey region, we cannot make a conclusive statement about the stability, as $\langle K_{\uparrow\downarrow} \rangle/\mathcal{L}d^{-1}$ has an almost linear behavior, as we observe from Fig.~\ref{fig:kpersol79}(a). This linear behavior would suggest an unstable regime and any small change in the numerical determination and the extrapolation could render the unstable behavior stable and vice versa.
It is also interesting to obtain some information about the stability of the system in the thermodynamic limit. Thus, we performed the above procedure for different system sizes and we performed an extrapolation to $\mathcal{L}\to\infty$, as exemplified in the inset of Fig.~\ref{fig:kpersol79}(b) for $J_{\uparrow\downarrow}=62.5Jd^{-2}$. The values of $\langle K_{\uparrow\downarrow} \rangle/\mathcal{L}d^{-1}$ in the continuum limit have been plotted in Fig.~\ref{fig:kpersol79}(b). We observe that for $J_{\uparrow\downarrow}\gtrsim 18.75Jd^{-2}$ the self-consistent solutions remain stable also in the thermodynamic limit.

The non-trivial stable solutions obtained in the continuum limit are plotted in Fig.~\ref{fig:kpersol}, for multiple system sizes. All the nontrivial stable solutions have a finite occupation of the cavity field, independent of the discretization or system size. In the following we will concentrate on the states which are stable both in the continuum and thermodynamic limit. Our results suggest a sudden onset of the occupation of the cavity mode above a certain threshold of the pump strength.

\subsection{\label{sec:meissner}Characterization of the Meissner superfluid steady state in the homogeneous system}

\begin{figure}[hbtp]
\centering
\includegraphics[width=.5\textwidth]{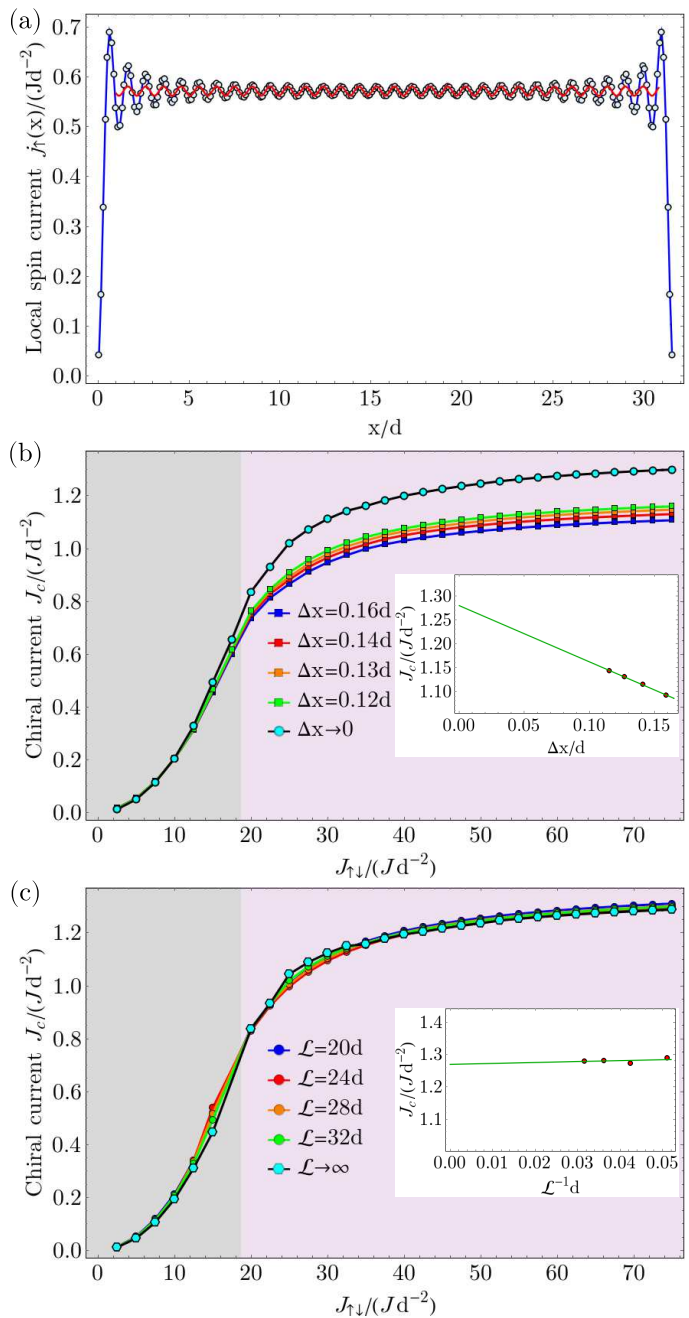}
\caption{(a) The local current $j_\uparrow(x)$ for the parameters $l_\varphi=0.8d$, $U=32.5Jd^{-1}$, $V=30Jd^{-1}$, $V_{trap}=0$, $\mathcal{L}=32d$, $J_{\uparrow\downarrow}=62.5Jd^{-2}$ and $\Delta x=0.12d$. The red curve represents the fit of the central part of the system, with the function $J_c/2+a \cos(b x+\phi)$. (b) The chiral current $J_c$ for the same parameters as a function of $J_{\uparrow\downarrow}$ for different values of the discretization and the extrapolation to the continuum limit. In the inset the extrapolation is shown for $J_{\uparrow\downarrow}=62.5Jd^{-2}$. (c) The chiral current in the continuum limit for different system sizes and extrapolated to the thermodynamic limit. The extrapolation for $J_{\uparrow\downarrow}=62.5Jd^{-2}$ is shown in the inset.
 }
\label{fig:jch}
\end{figure}

\begin{figure}[hbtp]
\centering
\includegraphics[width=.5\textwidth]{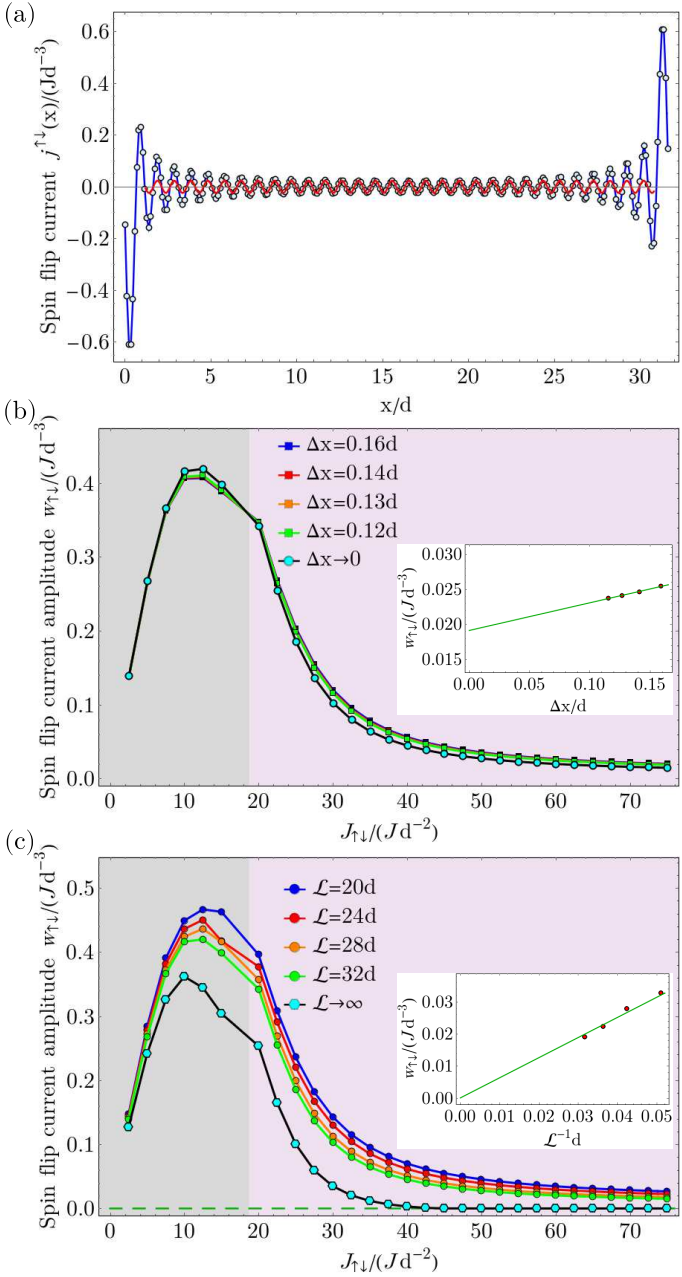}
\caption{(a) The local current $j^{\uparrow\downarrow}(x)$ for the parameters $l_\varphi=0.8d$, $U=32.5Jd^{-1}$, $V=30Jd^{-1}$, $V_{trap}=0$, $\mathcal{L}=32d$, $J_{\uparrow\downarrow}=62.5Jd^{-2}$ and $\Delta x=0.12d$. The red line represents the fit $w_{\uparrow\downarrow}\cos(a x+\phi)$ in the center of the system. (b) The amplitude of the oscillations in the spin flip current, $w_{\uparrow\downarrow}$, for the same parameters as a function of $J_{\uparrow\downarrow}$ for different values of the discretization and the extrapolation to the continuum limit. In the inset the extrapolation is shown for $J_{\uparrow\downarrow}=62.5Jd^{-2}$. (c) $w_{\uparrow\downarrow}$ in the continuum limit for different system sizes and the extrapolation to the thermodynamic limit. In the inset the extrapolation is shown for $J_{\uparrow\downarrow}=62.5Jd^{-2}$. In order to obtain an estimation of the spin flip current we enforce that it maintains a non-negative value in the $1/\mathcal{L}\to 0$ limit.
 }
\label{fig:jrung}
\end{figure}

\begin{figure}[hbtp]
\centering
\includegraphics[width=.5\textwidth]{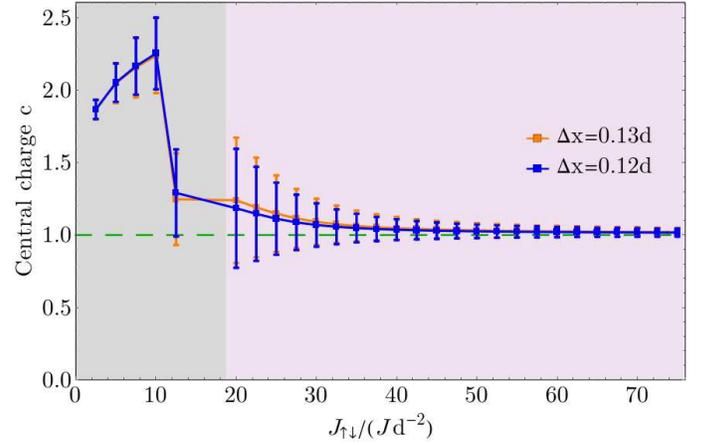}
\caption{The central charge for the parameters $l_\varphi=0.8d$, $U=32.5d^{-1}J$, $V=30Jd^{-1}$, $V_{trap}=0$, $\mathcal{L}=32d$ and multiple values of $\Delta x$, as a function of $J_{\uparrow\downarrow}$. The central charge is extracted fitting the scaling of the entanglement entropy. The errorbars represent the fit error. The dashed horizontal line indicates the constant value $1$.
 }
\label{fig:cc}
\end{figure}

In this section we show that the stable steady states, for $J_{\uparrow\downarrow}\gtrsim 18.75Jd^{-2}$, correspond to Meissner superfluid states. We focus on the behavior of the observables introduced in Sec.~\ref{sec:obs}. First we observe that all stable solutions, $J_{\uparrow\downarrow}\gtrsim 18.75Jd^{-2}$, have a finite value of the chiral current, shown in Fig.~\ref{fig:jch}. In Fig.~\ref{fig:jch}(a) we show the local current along the wire for one spin state, $j_\uparrow(x)$. In order to exemplify the extraction of the value of the chiral current we fit the oscillations due to the boundaries in the center part of the system for each leg, with the function $J_c/2+a \cos(b x+\phi)$. Extracting the chiral current as a function of $J_{\uparrow\downarrow}$ for different values of $\Delta x$ (Fig.~\ref{fig:jch}(b)) we obtain its behavior in the continuum $\Delta x \to 0$. By going to a smaller discretization we obtain a larger value of the chiral current, thus, we expect that it has a finite value in the continuum. If we perform this for multiple sizes of the system, shown in Fig.~\ref{fig:jch}(c), we obtain consistent results, indicating that the finite chiral current survives also in the thermodynamic limit. As we increase $J_{\uparrow\downarrow}$ and go deeper into the Meissner phase, the chiral current seems to saturate.

The Meissner state is further characterized by balanced spin flip processes in the bulk of the system. In the discrete ladder representation this means that there are no currents on the rungs of the ladder. We show the behavior of this observable in Fig.~\ref{fig:jrung}. First the site-resolved spin flip current has been plotted in Fig.~\ref{fig:jrung}(a), together with the fit function $w_{\uparrow\downarrow}\cos(a x+\phi)$, from which we extract the amplitude of the oscillations, $w_{\uparrow\downarrow}$. The results presented in Fig.~\ref{fig:jrung} were computed by fitting the central  $1/5$ part of the system. This differs with at most $6\%$ compared to the case if we would have considered the central $2/5$ of the system, $x\in(\mathcal{L}/2-\mathcal{L}/5,\mathcal{L}/2+\mathcal{L}/5)$.
We observe that the amplitude of the spin flip current, $w_{\uparrow\downarrow}$, first increases with $J_{\uparrow\downarrow}$. It reaches a maximum for $J_{\uparrow\downarrow}\approx 12.5Jd^{-2}$ in the region where the stability is not clarified and then decreases for larger values of $J_{\uparrow\downarrow}$ (Fig.~\ref{fig:jrung}(b)). 
The finite spin flip currents would indicate that a vortex state might be present for low values of $J_{\uparrow\downarrow}$. A phase transition may occur in the effective model between the Meissner state and a vortex state, close to the stability threshold.
In Fig.~\ref{fig:jrung}(c), the amplitude of the spin flip current extrapolated to the continuum limit is represented as a function of $J_{\uparrow\downarrow}$ for different system sizes. In the thermodynamic limit the spin flip current vanishes deep in the Meissner phase, as expected. However, in the intermediate regime for $18.75Jd^{-2}\lesssim J_{\uparrow\downarrow}\lesssim 35Jd^{-2}$ the amplitude of the spin flip current is larger than zero in the thermodynamic limit. This might be due to finite size effects or the extrapolation being too rough in this regime.
The amplitude of the oscillations in the particle density also goes to zero for large values of $J_{\uparrow\downarrow}$, after taking both the continuum and thermodynamic limit (not shown).

\begin{figure}[hbtp]
\centering
\includegraphics[width=.5\textwidth]{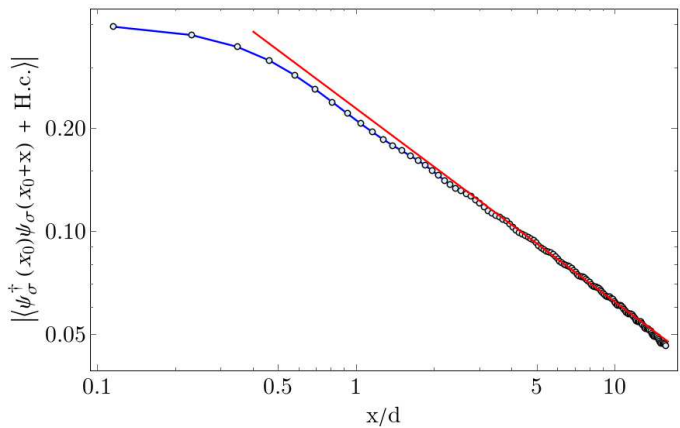}
\caption{The absolute value of the single particle correlations $\langle \psi^\dagger_{\sigma}(x_0) \psi_{\sigma}(x_0+x)+\text{H.c.} \rangle$  in a logarithmic plot for the parameters  $l_\varphi=0.8d$, $U=32.5d^{-1}J$, $V=30Jd^{-1}$, $V_{trap}=0$, $\mathcal{L}=32d$, $\Delta x=0.12d$ for $J_{\uparrow\downarrow}=62.5Jd^{-2}$, in the Meissner superfluid phase. The correlations show an algebraic decay with distance, which corresponds to the superfluid phase. The straight (red) line are fits of the function $\propto x^{-\alpha}$, where $\alpha=0.563\pm 0.002$.
 }
\label{fig:corr}
\end{figure}

Moreover, in the Meissner superfluid the central charge has the value, $c=1$. This agrees with our numerical data within the uncertainties as presented in Fig.~\ref{fig:cc}. Furthermore, we observe that for $J_{\uparrow\downarrow}\lesssim12.5Jd^{-2}$ the central charge has a value $c\approx 2$, which is consistent with the assumption of a vortex state. It can be seen that close to the stability threshold both the error bars due to the fit error and the difference between the values of the central charge computed for different discretizations are increasing. 

The superfluid nature of the stable stationary state solutions is in agreement with the algebraic decay of the single particle correlations, shown in Fig.~\ref{fig:corr} for $J_{\uparrow\downarrow}=62.5Jd^{-2}$. We mention that, in the weakly interacting limit, a Non-Luttinger liquid has been predicted \cite{PoZhou2014} at the critical point of the transition between the vortex and Meissner superfluids, with an exponential decay of the single particle correlations.

To summarize, we found that the dynamical stabilization of a Meissner superfluid state is possible, both in the continuum and thermodynamic limit.

\subsection{\label{sec:trap}The effect of the parabolic trap on the steady states}

\begin{figure}[hbtp]
\centering
\includegraphics[width=.5\textwidth]{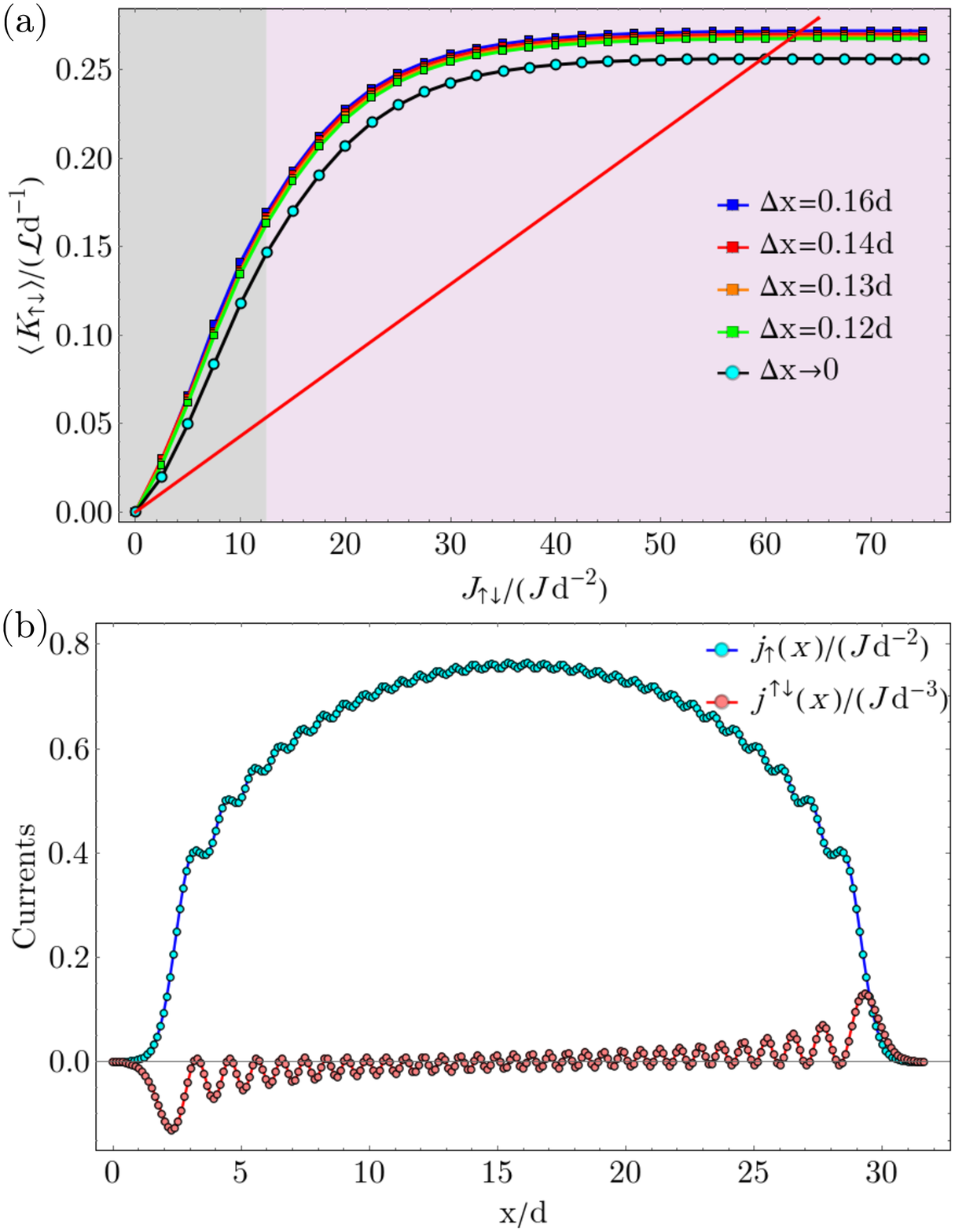}
\caption{(a) Graphical interpretation of the self-consistency condition for the parameters $N=32$, $l_\varphi=0.8d$, $U=32.5Jd^{-1}$, $V=30Jd^{-1}$ and $V_{trap}=5Jd^{-1}$. The expectation value $\langle K_{\uparrow\downarrow} \rangle/\mathcal{L}$ is represented for multiple values of $\Delta x$ and an extrapolation to the limit $\Delta x\to 0$. The straight (red) line represents the right-hand side of the self-consistency condition, which is a linear function with slope $\frac{J}{A\mathcal{L}d}$. The crossings of the two curves give the solutions of the self-consistency condition. (b) The local currents $j_\uparrow(x)$ and $j^{\uparrow\downarrow}(x)$  for the same parameters and $J_{\uparrow\downarrow}=62.5Jd^{-2}$ and $\Delta x=0.12d$. 
 }
\label{fig:trap}
\end{figure}

In this section we will investigate the effect of the harmonic trapping $H_{trap}$ on the dynamically organized steady states.

In the following, we will show that the Meissner state is stable also in the presence of a harmonic trap of strength $V_{trap}=5Jd^{-2}$, for $l_\varphi=0.8d$, $U=32.5Jd^{-1}$, $V=30Jd^{-1}$. 
Due to the varying density, a coexistence of different states is possible across the trap. However in determining the stability of the dynamically organized states all regions are important, as $\langle K_{\uparrow\downarrow} \rangle/\mathcal{L}$, which enters the self-consistency condition, is a global observable.
In the following, we will focus on the states that are present in the center of the trap, where the gradient of the trapping potential is small. 

In Fig.~\ref{fig:trap}(a), the expectation value $\langle K_{\uparrow\downarrow} \rangle/\mathcal{L}$ as a function of $J_{\uparrow\downarrow}$ is plotted in the trap. We obtain stable steady states for $J_{\uparrow\downarrow}\gtrsim 12.5Jd^{-2}$. In the following we focus on stationary state for $J_{\uparrow\downarrow}=62.5Jd^{-2}$, which is stable for all values of $\Delta x$ computed and in the extrapolation to the continuum limit. Due to the trapping potential, the large oscillations at the boundaries are partially suppressed, as we can observe from Fig.~\ref{fig:trap}(b), compared to the homogeneous case. We identify the state realized for these parameters in the center of the trap as a Meissner superfluid, due to its finite chiral current, small values of the spin flip current and the algebraic decay of correlations (not shown). 
The local spin current $j_\uparrow(x)$ has a maximum in the center of the trap (Fig.~\ref{fig:trap}(b)), different from the homogeneous case where $j_\uparrow(x)$ has a plateau-like behavior, see Fig.~\ref{fig:jch}(a). As $j_\uparrow(x)$ and $j^{\uparrow\downarrow}(x)$ are related via a continuity equation, at the edges of the trap the spin flip current has mostly negative values, for $x\lesssim 10d$, or positive values, for $x\gtrsim 22d$. This implies that the amplitude of the oscillations of $j^{\uparrow\downarrow}(x)$ is small only in the central region of the trap.
We note that the oscillations present in the spin flip currents are due to the finite size of the system and boundary effects. 

Thus, we found that also in the presence of a harmonic trapping potential the dynamical stabilization of the Meissner superfluid is possible.

\section{\label{sec:conclusion}Conclusions}

In this work we find a dynamically stabilized Meissner superfluid of bosonic atoms confined in a one-dimensional wire coupled to an optical cavity. The bosonic atoms are prepared in two hyperfine states coupled to each other via Raman transitions, which involve the creation or annihilation of a cavity photon and a position dependent momentum transfer. By this coupling a cavity mediated spin-orbit coupling is induced if a finite cavity occupation forms. A dissipative dynamics takes place due to the leaking of photons out of the cavity mirrors. Above a certain pump strength a nontrivial chiral state is realized as a steady state of the dissipative attractor dynamics. 

An experimental realization could  use \textsuperscript{87}Rb atoms in an optical cavity \cite{BaumannEsslinger2010, WolkeHemmerich2012,LeonardDonner2017} subjected to additional optical lattices which confine the atoms to one-dimensional structures. Previously, experiments of ultracold bosonic atoms in a cavity \cite{RitschEsslinger2013} observed the Dicke phase transition \cite{BaumannEsslinger2010, KlinderHemmerich2015, DomokosRitsch2002, DimerCarmichael2007, NagyDomokos2008, PiazzaZwerger2013} and an optical lattice has been added to investigate the influence of interaction \cite{KlinderHemmerichPRL2015, LandigEsslinger2016, HrubyEsslinger2018}.
 Typical cavity parameters are $\kappa= 2\pi \times 4.5~\text{kHz}$ and $g_0=2\pi \times 0.76~\text{MHz}$ \cite{WolkeHemmerich2012}. The cavity-induced spin-orbit coupling could be implemented using for example the two states from the \textsuperscript{87}Rb 5S\textsubscript{1/2} $F=1$ manifold, $\ket{\uparrow}\equiv\ket{F=1, m_F=0}$ and $\ket{\downarrow}\equiv\ket{F=1, m_F=-1}$ as used in Ref.~\cite{LinSpielman2011}. The scattering lengths of the two \textsuperscript{87}Rb states are $a_{\uparrow\uparrow}=100.86 a_B$ and $a_{\downarrow\downarrow}=100.4 a_B$ \cite{WideraBloch2006}, with $a_B$ the Bohr radius.
 
The nature of the steady state can be inferred by performing different measurements. A non-destructive measurement is the occupation of the cavity field by observing the leaking of the photons of the cavity. A finite occupation of the cavity field will identify the dynamically organized spin-orbit coupling and decide between the trivial and non-trivial steady states.  
For the state of the atomic component of the system, the measurement of the local density can distinguish between a Meissner and a vortex state. In the Meissner case a density plateau in the center part of the wire is present, in contrast to a vortex state which has density modulations. Furthermore, from the momentum distribution obtained in a time-of-flight experiment, one can check the dependence of the peak in the momentum distribution and the momentum transferred by the spin-orbit coupling. In Ref.~\cite{GreschnerVekua2016} by Greschner et al. (2016) the connection between the peak in the momentum distribution and the imprinted phase is made in the context of bosonic ladders.

\section*{\label{sec:acknowledgments}Acknowledgments}
We thank T.~Donner and T.~Esslinger for stimulating discussions. 
We acknowledge funding from the European Research Council (ERC) under the Horizon 2020 Research and Innovation Program, Grant Agreement No. 648166 (Phonton), and by the Deutsche Forschungsgemeinschaft (DFG, German Research Foundation) under FOR1807, Project No. 277625399 – TRR 185 Project B3 and CRC1238 Project No. 277146847 – Project C05. A.S. thanks the research council of Shahid Beheshti University, G.C.

\end{document}